\documentclass[aps,nofootinbib,superscriptaddress,twocolumn,prd,10pt]{revtex4-1}

\usepackage{graphicx}
\usepackage{amsmath,amssymb}
\usepackage{hyperref}
\usepackage{braket}
\usepackage{booktabs}
\usepackage{subfigure}
\usepackage{float}
\usepackage[dvipsnames]{xcolor}
\usepackage{mathrsfs}
\usepackage{comment}
\usepackage{bm}
\usepackage[normalem]{ulem}

\hypersetup{
    colorlinks=true,
    linkcolor=red,
    citecolor=blue,
}

\newcommand{\therm}{{\rm th}}

\newcommand{\zo}{{(0)}}
\newcommand{\fo}{{(1)}}
\newcommand{\so}{{(2)}}
\newcommand{\vev}[1]{ \left< {#1} \right> }

\begin{document}
\title{Spectral Distortion Anisotropy from Inflation for Primordial Black Holes}

\author{David Zegeye, Keisuke Inomata, Wayne Hu}
\affiliation{Kavli Institute for Cosmological Physics, Department of Astronomy \& Astrophysics, Enrico Fermi Institute, The University of Chicago, Chicago, IL 60637, USA}

\begin{abstract}
Single field inflationary models that seek to greatly enhance small scale power in order to form primordial black holes predict both a  squeezed bispectrum that is enhanced by this small scale power and a potentially detectable enhancement of CMB spectral distortions.   Despite this combination, spectral distortion anisotropy on CMB scales
remains small since the squeezed bispectrum represents an unobservable  modulation of the scale rather than local amplitude for the short wavelength acoustic power that dissipates and forms the $\mu$ spectral distortion.
The leading order amplitude effect comes from the local modulation of acoustic dissipation at the beginning of the $\mu$ epoch at the end of thermalization by a long wavelength mode that is correlated with CMB anisotropy itself.   Compensating factors from the suppression by the square of the ratio the comoving horizon at  thermalization to the smallest detectable primary CMB scales
($\sim 0.0005$)
and maximal allowed enhancement of $\mu$ ($\sim  5000$) leaves a signal in the $\mu T$ cross spectrum that is still well beyond the capabilities of PIXIE or LiteBIRD due to sensitivity and resolution while remaining much larger than in single field slow roll inflation and  potentially observable.   
\end{abstract}

\maketitle
\section{Introduction}

There has been much recent interest in primordial black holes (PBHs) from  an extremely large enhancement of small scale fluctuations during inflation 
(e.g.~\cite{Kawasaki:2016pql,Garcia-Bellido:2017mdw,Ezquiaga:2017fvi,Kannike:2017bxn,Germani:2017bcs,Ballesteros:2017fsr,Inomata:2017vxo,Hertzberg:2017dkh,Motohashi:2017kbs,Cheng:2018qof,Byrnes:2018txb,Passaglia:2018ixg,Drees:2019xpp,Carrilho:2019oqg,Passaglia:2019ueo,Ng:2021hll,Inomata:2021uqj,Inomata:2021tpx})
given their potential to explain dark matter
and the binary black hole mergers detected by 
LIGO-Virgo~\cite{Bird:2016dcv,Clesse:2016vqa,Sasaki:2016jop,Kashlinsky:2016sdv,Kashlinsky:2018mnu,Garcia-Bellido:2020pwq}. Moreover primordial fluctuations on scales much smaller than those probed by the cosmic microwave background (CMB) and large-scale structure are currently relatively poorly constrained
\cite{Josan:2009qn,Assadullahi:2009jc,Inomata:2018epa,Kawasaki:2021yek}.

Spectral distortions in the CMB are one way to constrain primordial power on small scales.   After the thermalization epoch where photon number changing processes in the plasma drop out of equilibrium, the energy dissipated in small scale acoustic waves leaves observable distortions in the frequency spectrum \cite{1970Ap&SS...7....3S,1991ApJ...371...14D,Hu:1994bz,Chluba:2011hw,Chluba:2013wsa}.  The amplitude of these acoustic waves is itself enhanced if the PBH scale is not much smaller than the dissipation scale at the end of thermalization.   The current bounds for chemical potential or $\mu$ distortions from COBE-FIRAS already place strong bounds on
such models and rule out PBHs as a significant fraction of the dark matter between $10^4-10^{13} M_\odot$~\cite{Nakama:2017xvq,Carr:2020gox}.

While current spectral distortion limits can be greatly improved with future space-based spectrometers~\cite{Chluba:2019nxa,LiteBIRD:2020khw}, an absolute measurement is limited by contamination from foregrounds and systematics in addition to instrument sensitivity.   Anisotropy in spectral distortions, if they are large and correlated with CMB anisotropy in temperature and polarization, provides a  promising complementary approach that uses cross correlation and differential measurement to mitigate these issues
\cite{Pajer:2012vz,Ganc:2012ae,Pajer:2012qep,Chluba:2016aln,Cabass:2018jgj,Remazeilles:2018kqd,Remazeilles:2021adt}.   

These correlations can arise if the amplitude of small scale power is modulated by long-wavelength fluctuations due to the squeezed bispectrum.  However in single-field slow roll inflation, the squeezed bispectrum obeys the Maldacena consistency relation \cite{Maldacena:2002vr}, where long wavelength curvature fluctuations modulate the scale and not the amplitude of the
 small scale power spectrum.  Moreover this scale modulation is unobservable locally since a coordinate system established using clocks and rulers locally cannot reference the global coordinate system in which the long-wavelength mode is embedded \cite{Pajer:2013ana}. 
 Since spectral distortions depend only on the amount of power dissipated and not the globally referenced comoving scale from which it originates, single-field slow roll inflation does not produce spectral distortion anisotropy at leading order \cite{Pajer:2012vz,Cabass:2018jgj}.
 
 Recently it has been suggested that single field PBH models may 
 violate the conditions that make the spectral distortion anisotropy vanish at leading order in the squeezed bispectrum \cite{Ozsoy:2021pws,Ozsoy:2021qrg}.  All such PBH models must violate the slow roll
 assumption in order to enhance small scale power sufficiently rapidly \cite{Motohashi:2017kbs}.   Proposed PBH models typically have a period of non-attractor behavior which causes a violation of the
 Maldacena consistency relation for long wavelength modes that exit the horizon sufficiently close to or in the non-attractor phase \cite{Namjoo:2012aa,Martin:2012pe,Cai:2018dkf,Passaglia:2018ixg}.
 
 In this work we show that for scales relevant for correlation with CMB anisotropy, which are much, much larger than the horizon at the onset of the non-attractor phase, the Maldacena consistency relation holds in this limit and spectral distortion anisotropy is suppressed by the square of the ratio between the long wavelength scale and a characteristic short wavelength scale.  
 Nevertheless, these suppressed effects can still be much larger than 
 they are in slow roll inflation \cite{Cabass:2018jgj} and appear mainly due to the modulation of the dissipation scale at the end of thermalization by the long wavelength mode. 
 
 The outline of this paper is as follows.   In \S \ref{sec:USR}, we review non-attractor  ``ultra-slow roll" mechanism for enhancing small scale power during inflation and discuss the modulation of short-wavelength modes by long-wavelength modes in the form of the consistency relation.   For the relevant long-wavelength scales we show that the usual Maldacena consistency relation holds and, in 
 Appendix \ref{sec:secondorder}, we explicitly verify that once combined with its impact on short-wavelength acoustic evolution, the zeroth order effect of the long-wavelength mode is a dilation of scales that is unobservable locally.   In \S \ref{sec:acoustic} and \S \ref{sec:thermalization}, we calculate the
 leading order effect of the long wavelength density fluctuation on the local amplitude of
 acoustic oscillations and their dissipation at the end of thermalization respectively.   We discuss the implications of this modulation on spectral distortion anisotropy in \S \ref{sec:thermalization}.   In \S \ref{sec:snr} we assess the prospects for detecting this signal and discuss these results in \S \ref{sec:discussion}.

\section{USR Consistency Relation}
\label{sec:USR}

A common aspect of many inflationary models that enhance the small scale curvature power spectrum by the orders of magnitude that would be required to later form PBHs in the radiation dominated epoch is a transient period of so-called ``ultra slow-roll" (USR) \cite{Kinney:2005vj}.  During USR, the inflaton $\phi$ rolls faster than can be sustained by the slope of its potential $V(\phi)$,
\begin{equation}
\left| \frac{d\phi}{dN} \right| \gg \left| \frac{V'(\phi)}{H^2} \right| ,
\end{equation}
where $H=d\ln a/dt= dN/dt$ is the Hubble parameter and $N$ is the efold. 
This excess kinetic energy can arise for example from a very flat potential around an inflection point (e.g.\ \cite{Garcia-Bellido:2017mdw}) or from a sudden increase in kinetic energy due to a downward feature in the potential
\cite{Inomata:2021uqj}. In either case, the excess kinetic energy of the field then redshifts away as $(d\phi/dN)^2 \propto a^{-6}$ and the curvature fluctuation in unitary gauge grows as $\zeta = -\delta\phi/(d\phi/dN) \propto a^{3}$ once it crosses the horizon.  

Modes that were already outside the horizon at the beginning of the USR phase also experience growth to the extent that they have not completely frozen out.   Following the treatment in 
Ref.~\cite{Passaglia:2018ixg}, during the preceding  slow roll (SR) phase  superhorizon
curvature fluctuations evolve to a constant as $d\ln\zeta/dN \approx - (k/aH)^2$.  Once USR begins $d\zeta/dN \propto a^3$ and so the value of $(k/aH)_{\rm USR}^2$, where the subscript denotes the beginning of the USR epoch, determines the growth of these modes.   Given the opposite sign of $d\ln\zeta/dN$ during SR, there is a mode for which this growth is just sufficient to overcome the initially constant SR piece and cause a near zero crossing in the curvature power spectrum at $k=k_{\rm dip}$ where
\begin{equation}
\left(\frac{k_{\rm dip}}{a H}\right)^2_{\rm USR} \sim
e^{-3 \Delta N_{\rm USR}},
\end{equation}
and $\Delta N_{\rm USR}$ is the number of efolds in the USR phase.  For example for the $\sim 10^{7}$ enhancement of power typical to PBH models 
$\Delta N_{\rm USR} \approx \ln (10^7)/6$ and so $k_{\rm dip} \sim 10^{-2} (aH)_{\rm USR}$.
Note that these scalings also imply that
the enhancement in power for $k_{\rm dip} < k \lesssim (aH)_{\rm USR}$ scales as $k^4$ \cite{Byrnes:2018txb}.  
Thus the curvature power spectrum for $k\lesssim
(aH)_{\rm USR}$ can be parameterized as
\begin{equation}
 \label{eq:PBHspectrum}
 \Delta^{2}_\zeta(k)\equiv \frac{k^3 P_\zeta(k)}{2\pi^2} = \Delta^{2}_{{\rm SR}}(k) \left[1-\left(\frac{k}{k_{\rm dip}}\right)^2\right ]^2, 
\end{equation}
where $\Delta^{2}_{\rm SR}$ is the SR power spectrum which we take for simplicity to be a pure power law
\begin{equation}
\Delta^{2}_{\rm SR}(k)= A_s \left( \frac{k}{0.05\, {\rm Mpc}^{-1} } \right)^{n_s-1},
\end{equation}
with $A_s = 2.1\times 10^{-9}$ and $n_s=0.965$
consistent with CMB measurements~\cite{Aghanim:2018eyx}.  Note that this form can be derived more rigorously in specific models and it holds to leading order in the downward step 
PBH model (see \cite{Inomata:2021tpx} Eq.~(39)).
For $k\gtrsim (aH)_{\rm USR}$ the result becomes dependent on the specific  PBH model but for the modes relevant for spectral distortions that satisfy current observational constraints Eq.~(\ref{eq:PBHspectrum}) suffices.

\begin{figure}[t!]
\begin{center}
\includegraphics[width=1\columnwidth]{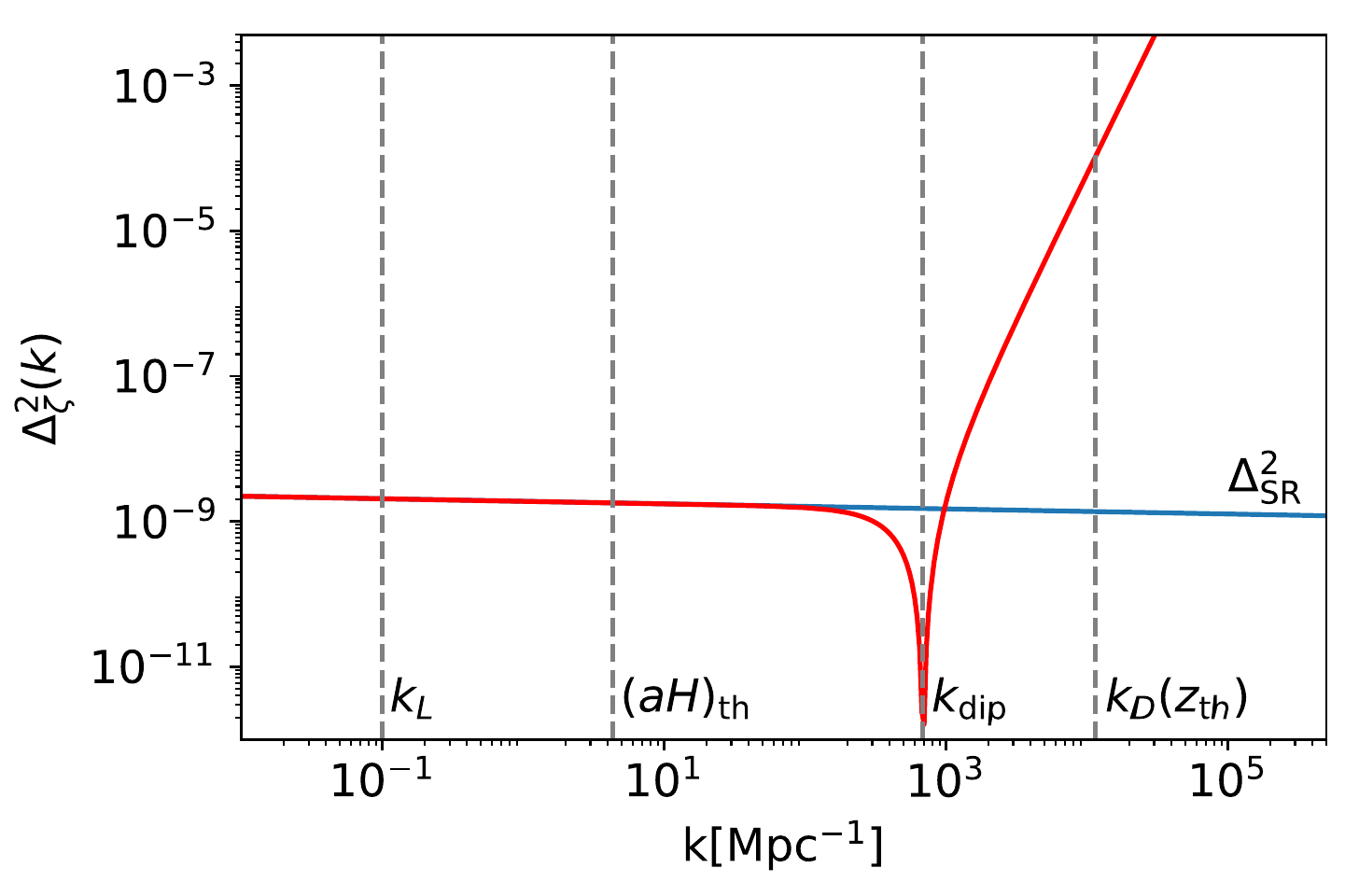}
\end{center}
\caption{\label{fig:powerspectrum}
The PBH curvature power spectrum $\Delta^2_\zeta(k)$ for a USR enhancement  starting at $k_{\rm dip}=681$ Mpc$^{-1}$ which provides the largest spectral distortion still allowed (see Fig.~\ref{fig:mukdip}).   Vertical lines
show $k_L=0.1$ Mpc$^{-1}$ (of order the largest $k$ mode probed by the CMB), $(aH)_{\rm th}$ the comoving size of the horizon at the end of thermalization,
$k_{\rm dip}$, and the dissipation wavenumber at the end of thermalization
$k_D(z_{\rm th})$.  The  slow roll power law extrapolation $\Delta_{\rm SR}^2$ is also shown for reference.}
\end{figure}

In Fig.~\ref{fig:powerspectrum}, we show the power spectrum (\ref{eq:PBHspectrum}) for $k_{\rm dip}=681$ Mpc$^{-1}$, which we shall see below in Fig.~\ref{fig:mukdip} is the largest scale allowed by current constraints on spectral distortions which arise from the power near the dissipation scale at end of thermalization $k_D(z_\therm)\approx 10^{4}$ Mpc$^{-1}$.  Notice the large hierarchy of scales between these values and the smallest scale accessible to measurements of the primary CMB anisotropy $k_L \sim 0.1$\, Mpc$^{-1}$ 
due to the dissipation scale at recombination.

The wavenumber $k_{\rm dip}$ separates two very different regimes for the impact of long wavelength fluctuations on much shorter wavelength power or equivalently the squeezed bispectrum.   For a long wavelength
$k_L\ll k_{\rm dip}$ the squeezed bispectrum obeys the usual Maldacena consistency relation \cite{Maldacena:2002vr}
\begin{eqnarray}
\lim_{k_L/k_S \rightarrow 0} 
B_\zeta(k_L,k_S,k_S) &=& -\left[\frac{ d\ln \Delta^2_\zeta(k_S)}{d\ln k_S}+ {\cal O}\left(\frac{k_L}{k_{\rm dip}}\right)^2\right] \nonumber\\
&&\times P_\zeta(k_L) P_\zeta(k_S),
\label{eq:consistency}
\end{eqnarray}
which can be demonstrated by explicit calculations using the in-in or $\delta N$ formalisms \cite{Passaglia:2018ixg}.  The interpretation of this relation is that the {\it scale} of the short wavelength modes are 
 dilated  by the nearly constant long wavelength curvature perturbation, which acts as a spatial 
fluctuation in the local scale factor
\begin{eqnarray}
\Delta_\zeta^2(k_S,{\bm y}) & \approx & \Delta_\zeta^2
(k_S (1-\zeta_L({\bm y})) 
\nonumber\\
\label{eq:modulation}
& \approx &
 \Delta_\zeta^2
(k_S) \left(1 - \frac{ d\ln \Delta^2_\zeta(k_S)}{d\ln k_S} \zeta_L({\bm y})\right),
\end{eqnarray}
where local spatial variations are denoted by the comoving coordinate ${\bm y}$.
Notice that the amplitude of the power spectrum does not actually change locally, just the comoving scale that the power is associated with.   
Physically, the long wavelength mode just changes the efold at which the short wavelength mode enters into USR, not any of the dynamics due to the USR growth.   As shown in Ref.~\cite{Cabass:2018jgj} this change in scale cannot modulate spectral distortions, which depend only on the amplitude of the power dissipated, not the scale.  In the Appendix we explicitly verify  that a constant $\zeta_L$ generates a locally unobservable dilation for both the  primordial non-Gaussianity and the subsequent dynamics of short wavelength modes to second order in perturbations.

For $k_L\gtrsim k_{\rm dip}$, the leading effect of the long wavelength mode is not simply a dilation of scales and hence the consistency relation (\ref{eq:consistency}) no longer holds \cite{Namjoo:2012aa,Martin:2012pe}.
Physically, the inflaton field fluctuation of the long wavelength mode changes the number of efolds that the short wavelength mode experiences USR
since there is no longer an attractor solution that makes it equivalent to a shift along the background phase space trajectory
\cite{Passaglia:2018ixg}.  For example, a constant backwards fluctuation of the field means that locally the short wavelength modes see more efolds of USR growth and so the actual amplitude of the small scale power spectrum is modulated by the long wavelength mode.   This then would produce a spatial modulation in spectral distortions.   

We shall see that for modes relevant for correlation with CMB anisotropy and models that satisfy spectral distortion constraints, $k_L \ll k_{\rm dip}$. 
Even in this regime, there is always some dynamical effect of short wavelength growth in the long wavelength field to the extent that $\zeta_L \ne$\ const.  As we have seen, here  the evolving part of $\zeta_L$ during USR is suppressed by $k_L^2/k_{\rm dip}^2$. This expectation is consistent with
an explicit calculation of the leading order correction the Maldacena consistency relation given in  
Eq.~(\ref{eq:consistency}) in an inflection point PBH model \cite{Passaglia:2018ixg}.
However since again the relevant $k_L \ll k_{\rm dip}$ this correction produces a very small $\mu T$ correlation (see Fig.~\ref{fig:powerspectrum} and \S \ref{sec:mu}).

\section{Local Expansion} 
\label{sec:local}

After inflation, there are $(k_L/aH)^2$ suppressed   modulations of short wavelength physics by the long wavelength curvature fluctuations $\zeta_L$, analogous to the corrections to the primordial bispectrum from $k_{\rm dip} \propto (aH)_{\rm USR}$
in the previous section.
To the local observer, the long wavelength mode appears as a change to the background cosmology induced by the density perturbation that it carries.  Since synchronous observers are freely falling test particles, we can absorb these synchronous gauge adiabatic density fluctuations $\delta_L$ \cite{Hu:2016ssz,Hu:2016wfa},
\begin{equation}
\delta_L = \frac{1}{3} \left( \frac{k}{aH} \right)^2  \zeta_L \propto a^2
\end{equation}
during radiation domination
into a new background or ``separate universe".  

Following the construction of the separate universe for the late time growth of structure \cite{Sirko:2005uz,Li:2014sga}, we can define a local background density $\bar\rho_L$ as
\begin{equation}
\bar \rho (1+ \delta_L) = \bar\rho_L,
\end{equation}
which implies that the local observer sees a scale factor  $a_L$ that is related to the global scale factor $a$ at equal times as
\begin{equation}
a_L \approx  a (1- \delta_L/4),
\end{equation}
where we chosen the normalization such that $a_L \rightarrow a$ at early times.
Notice that this normalization removes the dilation effect of $\zeta_L$ in
Eq.~(\ref{eq:modulation}) by measuring scales locally so that they coincide when $\delta_L \ll 1$ at the end of inflation.

By virtue of this normalization and conservation of particles, at the same numerical value of the scale factors (or efolds) from the end of inflation in the local and global universe, all particle number and energy densities are the same.   However  the scale factors do not coincide at the same time
 in the local and global universe.

To extract the cosmological parameters of the separate universe, we can express the expansion rate as a function of the local scale factor.
With the definition $H_L \equiv d\ln a_L/dt$ and the radiation dominated growth of $\delta_L \propto a^2$ we have at equal times
\begin{equation} 
H_L^2 \approx H^2 (1-\delta_L).
\end{equation}
In the global universe let us define a reference epoch $a_r$ in the radiation dominated limit where $H^2(a) = H_r^2 (a_r/a)^4$ so that we can express the local expansion rate in terms of the local scale factor as
\begin{equation}
\label{eq:HubbleL}
H_L^2 = H_r^2 \left( \frac{a_r}{a} \right)^4 (1-\delta_L)
\approx H_r^2 \left( \frac{a_r}{a_L} \right)^4 (1- 2\delta_L).
\end{equation}
In the local universe this takes the form of the Friedmann equation
with $\delta_L \propto a^2 \approx a_L^2$  playing the role of spatial curvature to linear order in $\delta_L$,
specifically
\begin{equation}
H_0^2 \Omega_{KL} = -2 {\delta_L}(a_r)  H_r^2 a_r^2
\end{equation}
so that  $\delta_L(a_r)$ at the reference epoch  defines the comoving curvature scale in units of the comoving Hubble scale at that epoch.  Finally it is useful to express
the local conformal time as
\begin{equation}
\eta_L = \int \frac{d\ln a_L}{ a_L H_L} \approx
\frac{a_L}{a_r} \frac{1}{ a_r H_r} \left(1 + \frac{\delta_L}{3} \right) .
\end{equation}
These relations now determine the modulation of
all short wavelength observables by a long wavelength curvature fluctuation $\zeta_L$ which  to leading order scales as $(k_L/aH)^2\zeta_L$.  The specific size of the modulation will then depend on the epoch at which it influences the short wavelength observable the most. 
We shall see that for $\mu$ distortions and PBH models, this is the end of the thermalization epoch.

\section{Modulated Acoustic Power}
\label{sec:acoustic}

The first step in understanding the local modulation of spectral distortions is to determine the impact of the long wavelength curvature perturbation $\zeta_L$ on
the amplitude of short wavelength acoustic oscillations in the CMB during radiation domination.   These oscillations then dissipate via diffusion damping leaving a spectral distortion after the 
thermalization epoch.   We shall see that since acoustic oscillations are generated at horizon crossing of the short wavelength mode $k_S$, the impact of the long wavelength mode occurs at horizon crossing of the short wavelength mode $aH = k_S$ and therefore scales as $(k_L/k_S)^2 \zeta_L$.  The impact of CMB scale wavenumbers $k_L$ is therefore highly suppressed for the modes that contribute to spectral distortions
(see Fig.~\ref{fig:powerspectrum}).

We can analytically understand this scaling in the simple case where the photons dominate the radiation density, i.e.\ neglecting the effect of neutrinos which only change the numerical factors and not the overall scaling.
In this case we can solve the perturbation equations in terms of the
continuity and Euler equations for the photon fluid under self gravity 
in the photon-dominated local universe as (see \cite{Hu:1996vq} Eq.~(10)) 
\begin{eqnarray}
 \Delta_\gamma' - \frac{y'}{y} \Delta_\gamma&=&-  \frac{4}{\sqrt{3}} \frac{ 
 1 -  {y''}/{y} + 2 ( {y'}/{y})^2 }{ 1 + {6}/{(f_K y^2)} } V_\gamma, \nonumber\\
V_\gamma' + \frac{y'}{y}V_{\gamma} &=& \left( 1- \frac{6}{f_K y^2} \right) \frac{\sqrt{3}}{4}  \Delta_\gamma,
\label{eqn:EOM}
\end{eqnarray}
where $\Delta_\gamma$ is the comoving gauge photon density perturbation, $y= (\Omega_\gamma H_0^2)^{-1/2} a_L k_S$, $' = d/dx$ with $x=k_S\eta_L/\sqrt{3}$, and
$f_K 
= 1+ 3\Omega_{KL} H_0^2/k_S^2$.  Here we have again made use of the fact that at the same value for the scale factor, the physical density of the photons is the same in the local and global universe.  
The synchronous and comoving gauge differ in their density perturbations outside the horizon due to radiation pressure so $\lim_{x\rightarrow 0} \Delta_\gamma \ne \delta_L$ but they do approach each other for $x\gg 1$ in the regime relevant for spectral distortions.

Without the curvature perturbation induced by the long wavelength mode, $x= y/\sqrt{3}$ during photon domination and the solution is analytic
\begin{equation}
\lim_{y\rightarrow {\sqrt{3}}x}\Delta_\gamma = 4 \left( \frac{\sin x}{x} -\cos x   \right) \zeta_S,
\label{eq:zeroth}
\end{equation}
from which we can read off the usual transfer function\footnote{This transfer function is reduced to
$-4/(1+ 4R_\nu/15)\sim -3.61$ when neutrinos with
$R_\nu = \rho_\nu/(\rho_\nu+\rho_\gamma)$ are included \cite{Hu:1995en}}
for acoustic oscillations $-4\cos x$ at $x\gg 1$

We can now solve for the leading order correction from the small curvature induced by the long wavelength mode.  Since we neglect neutrinos $H_r^2 = \Omega_\gamma H_0^2 a_r^{-4}$ and we have
\begin{equation}
y \approx {\sqrt{3}} x \left(1-\frac{\delta_L}{3} \right) 
\approx {\sqrt{3}} x \left(1 - \frac{x^2}{{3}}
\alpha \right),
\label{eq:y_app}
\end{equation}
so that Eq.~(\ref{eqn:EOM}) becomes to leading order:
\begin{equation}
\Delta_\gamma'' +
\left( 1- \frac{2}{x^2} -\frac{4}{3} \alpha  \right)\Delta_\gamma =0
\label{eqn:FirstOrderEOM}
\end{equation} 
with 
\begin{equation}
\alpha \equiv \left( \frac{k_L}{k_S} \right)^2 \zeta_L
\end{equation}
and $f_K = 1- 2 \alpha $
constant in time.

Notice that at $x\gg 1$ this takes the form of an oscillator equation
with a perturbed constant frequency
\begin{equation}
\tilde x = \left( 1- \frac{2}{3}\alpha \right)x .
\end{equation}
We can now iterate to solve Eq.~(\ref{eqn:FirstOrderEOM})
to first order in $\alpha$ using the zeroth order
solution (\ref{eq:zeroth}) to determine $\alpha \Delta_\gamma$ as an external source,
\begin{equation}
\Delta_\gamma \approx 4\zeta_S \left(1+\frac{4}{3}\alpha \right) \left(   \frac{\sin \tilde x }{\tilde x} -\cos \tilde x 
  \right).
  \label{eq:delta_gamma_mod}
\end{equation}
We can explicitly verify that this form solves 
(\ref{eqn:FirstOrderEOM}) and satisfies $\Delta_\gamma \approx 4\zeta_S x^2/3$ in $x\ll1$ to linear order in $\alpha$.

Therefore there is an $\alpha=(k_L/k_S)^2\zeta_L$  change in the amplitude and frequency of the acoustic wave.
Since $k_L/k_S \ll 1$, this is a large suppression factor and produces a negligible change in the local spectral distortion once the acoustic modes have dissipated.  We can therefore hereafter assume that the power in acoustic modes at $k_S$ is effectively the same in the global and local universe at the same value of the scale factors.

\section{Modulated Thermalization}
\label{sec:thermalization}

We can now compute the local dissipation of energy from  the acoustic waves into $\mu$ spectral distortions in the separate universe.  Changes in  thermalization due to the local background induce  larger local variations in $\mu$ for PBH models, than the primordial effect from $k_{\rm dip}$ or the acoustic growth, since we shall see they scale with the comoving horizon size at the end of thermalization
$(k_L/a H)_\therm^2 \zeta_L$.

Let us now see how thermalization is altered by the long-wavelength modulation in the local universe.
The thermalization rate for the joint action of double Compton scattering
$e^{-}+\gamma \leftrightarrow e^{-}+2\gamma$, which changes photon number, and Compton 
scattering  $e^{-}+\gamma \leftrightarrow e^{-}+\gamma$,
which redistributes energy, to establish a blackbody scales as (see e.g. \cite{1982A&A...107...39D,Hu:1992dc})
\begin{equation}
\Gamma_\therm \propto T^{3/2} n_e \propto a_L^{-9/2},
\end{equation}
where $n_e$ is the free electron density and $T$ is the plasma temperature and recall that at the same numerical value of the scale factors
all particle densities are the same in the local and global universe. 

Let us define the thermalization time by the condition
\begin{equation}
\int dt \Gamma_\therm  = \int_{\ln a_\therm}^0 d\ln a_L \frac{\Gamma_\therm}{H_L} = 1.
\end{equation}
Using Eq.~(\ref{eq:HubbleL}), the change in the efold of thermalization  is given by
\begin{equation}
\Delta \ln a_\therm = {2}\delta_L(a_{\therm})
\label{eq:athermL}
\end{equation}
such that in an overdensity, thermalization continues to a later efold due to a slower expansion rate.

The energy in acoustic waves is dissipated when the photon diffusion length crosses the wavelength.  If this occurs after the thermalization epoch but before the Compton scattering becomes inefficient at redistributing energy, then this energy is transferred into a $\mu$ spectral distortion (see Eq.~(\ref{eq:derive}) below).
The diffusion wavenumber in the radiation epoch 
is given by
\begin{equation}
k_D^{-2} \propto \int \frac{d\ln a}{a H} 
\frac{1}{n_e\sigma_T a},
\end{equation}
where $\sigma_T$ is the Thomson cross section
(see Eq.~(\ref{eq:diffusion}) below for the general expression).
Again let us make use of the fact that the particle densities are the same in the local and global universe at the same efold.
We then get the change in the diffusion wavenumber as
\begin{eqnarray}
\label{eq:dampingshift}
\Delta\ln k_D(a_\therm) &=& -\frac{3}{2}\Delta \ln a_\therm - \frac{3}{10} \delta_L(a_{\therm}) \\
&=& - \frac{33}{10}\delta_L(a_{\therm}) =
-\frac{11}{10} \left( \frac{k_L}{aH} \right)_\therm^2 \zeta_L,  \nonumber
\end{eqnarray}
where recall that in the global universe
$k_D \propto a^{-3/2}$.
Therefore in an overdensity the diffusion wavenumber decreases.
The net result is that the maximal wavenumber that dissipates into 
$\mu$ distortions given the same small scale power spectrum
in local coordinates as global coordinates is modulated by
$(k_L/aH)_\therm^2 \zeta_L$.   Since the power spectrum is strongly blue tilted, it is this modulation that changes the local value of $\mu$.
This is in contrast to single field slow roll inflation where the larger horizon size at the end of the 
$\mu$ epoch makes the modulation 
at that time the dominant effect
\cite{Cabass:2018jgj}.

\begin{figure}[t!]
\begin{center}
\includegraphics[width=1\columnwidth]{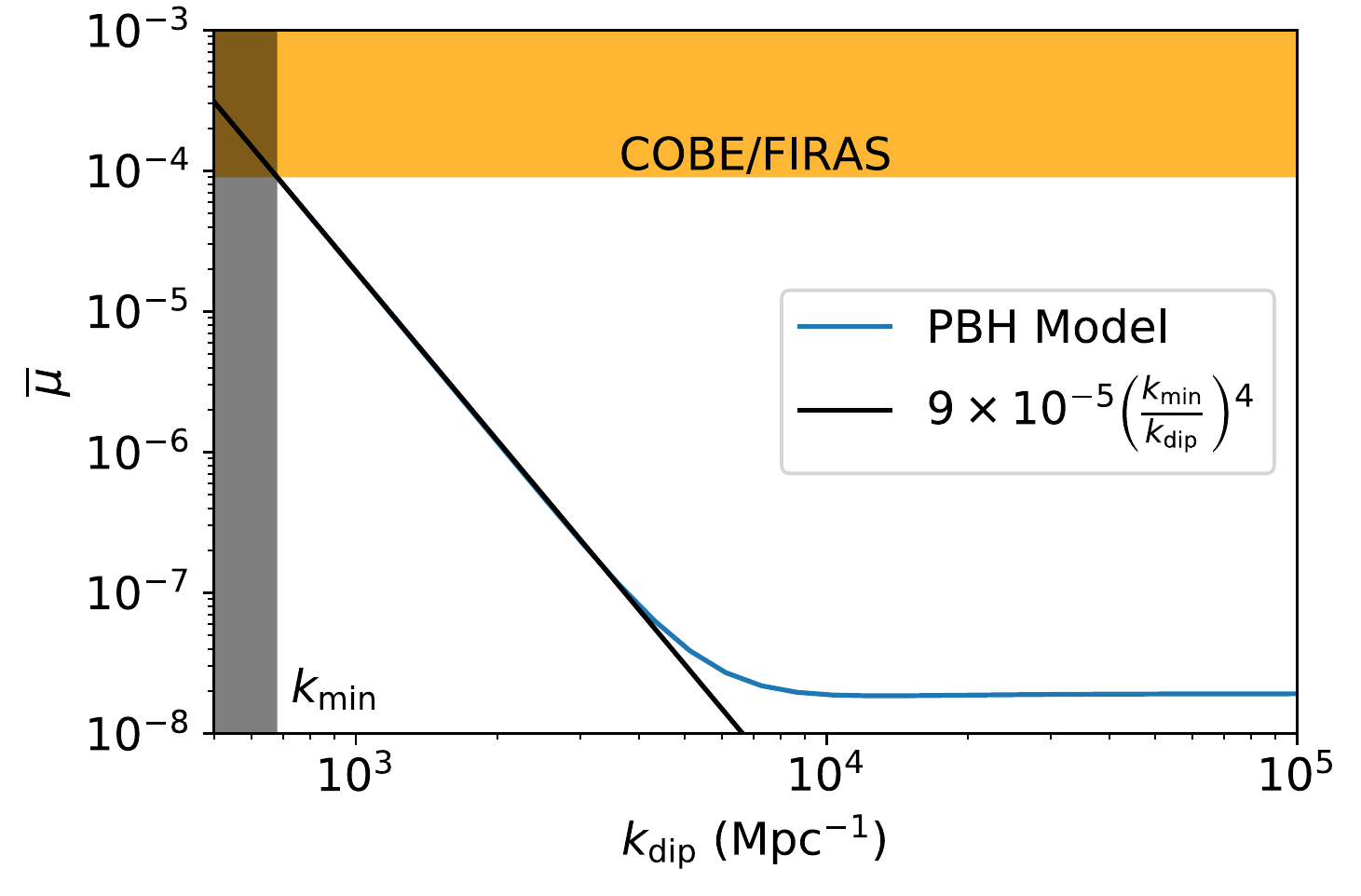}
\end{center}
\caption{\label{fig:mukdip}  The average spectral distortion $\bar\mu$ for various PBH enhancement scales $k_{\rm dip}$ (see Eq.~(\ref{eq:PBHspectrum})). Values of $k_{\rm dip} < k_{\rm min}$ are excluded by the FIRAS bound of $\bar \mu \leq 9 \times 10^{-5}$ whereas allowed values that lead to enhancement over the slow roll prediction of $\bar\mu \approx 2\times 10^{-8}$ follow $\bar\mu \propto k_{\rm dip}^{-4}$.
}
\end{figure}

\section{Spectral Distortion Anisotropy} 
\label{sec:mu}

With the local change in the thermalization and dissipation scales due to the long wavelength curvature perturbation as calculated in the previous section, we can now determine how it modulates the $\mu$ spectral distortion.   Because long wavelength curvature perturbations also generate CMB anisotropy, this leads to a potentially observable $\mu T$ correlation in PBH models.

First we calculate the average spectral distortion in the global universe with the PBH power spectrum of Eq.~(\ref{eq:PBHspectrum}).
Following \cite{Cabass:2018jgj,Chluba:2016aln}, 
\begin{equation}
    \bar \mu =\int d \ln k \Delta_\zeta^{2}(k) W(k),
\end{equation}
where to good approximation 
\begin{eqnarray}
    W(k)&\approx&-4.54 k^{2}  \int_{0}^{\infty}{d}z  \frac{d k_D^{-2}}{dz}  \mathcal{J}_{\mu}(z)\mathrm{e}^{-{2k^{2}}/{k_D^{2}(z)}}.
\end{eqnarray}
Here the diffusion wavenumber $k_D$ is given by
\begin{equation}
\label{eq:diffusion}
k_D^{-2}(z) = \frac{1}{6} \int_z^\infty \frac{dz}{H}
\frac{1}{n_e \sigma_T a} \frac{R^2 + 16(1+R)/15}{(1+R)^2},
\end{equation}
where $R = 3\rho_b/4\rho_\gamma$.
In the radiation dominated epoch and with the best fit $\Lambda$CDM parameters 
\begin{equation}
k_D \approx 4.05 \times 10^{-6} (1+z)^{3/2} \,{\rm Mpc}^{-1}.
\end{equation}
The Green function $\mathcal{J}_\mu$ for $\mu$ distortions is well approximated by \cite{Chluba:2013wsa}
\begin{eqnarray}
\label{eq:derive}
    \mathcal{J}_{\mu}\left(z\right) &\approx&  \left[1-\exp \left(-\left[\frac{1+z}{5.8 \times 10^{4}}\right]^{1.88}\right)\right] \Theta(z-z_{\rm rec}) \nonumber\\
    &&\times e^{-(z/z_{\therm})^{5/2} },
\end{eqnarray}
where we have included a $\Theta$ step function at recombination $z_{\rm rec}$ since below this redshift there are no acoustic waves to dissipate.
Here $z_\therm\approx 2\times 10^6$ is the thermalization redshift in the global universe and the quantity in brackets of Eq.~(\ref{eq:derive}) determines the gradual transition from the $\mu$ epoch to the Compton $y$ epoch.   Notice that $\bar\mu$ receives contributions from the initial inflationary power spectrum at wavenumber $k$ mainly when it crosses $k_D$ but these contributions are sharply cut off by the thermalization process above $k_D(z_\therm)$.  

In Fig.~\ref{fig:mukdip} we show $\bar\mu$ as a function of $k_{\rm dip}$ in Eq.~(\ref{eq:PBHspectrum}). 
The COBE-FIRAS constraint
$\bar\mu<9\times 10^{-5}$ (95\% CL)
\cite{Fixsen:1996nj} places a limit of
\begin{equation}
k_{\rm dip}>k_{\rm min} \approx 
681\, {\rm Mpc}^{-1}
\end{equation}
and 
\begin{equation}
\bar \mu \approx 9\times 10^{-5} \left(\frac{k_{\rm min}}{ k_{\rm dip} } \right)^{4} 
\label{eq:mubarscaling}
\end{equation} 
for models with  $k_{\rm min} < k_{\rm dip} \lesssim 5000$\ Mpc$^{-1}$ whereas for even larger $k_{\rm dip}$, $\bar\mu$ asymptotes to its  slow-roll value of $\bar\mu \approx 2\times 10^{-8}$ 
because enhanced scales have already dissipated by the thermalization epoch.  Notice that even this smallest allowed value {of $k_{\text{dip}}$} is much greater than the comoving horizon at the end of thermalization $(aH)_\therm \approx 4.3 $ Mpc$^{-1}$. 

The modulation of $k_D(z_\therm)$ due to the long wavelength mode therefore modulates the local value of $\mu$ from its background value $\bar \mu$.
Given that the PBH power spectrum rises as approximately
$\Delta_\zeta^2 \propto k^{{4 - (1-n_s)}}$ in the region which can enhance $\mu$,\footnote{For consistency with Eq.~(\ref{eq:PBHspectrum}) we have retained $1-n_s$ here but note that any actual ${\cal O}(1-n_s)$ correction to the local slope will depend on the details of the model.} we get the fractional change in the local
value of $\mu$ as 
\begin{equation}
\delta \ln \mu = [4-(1-n_s)]\Delta\ln k_D \equiv b_\therm \left( \frac{k_L}{a H}\right)_\therm^2 \zeta_L ,
\end{equation}
where from Eq.~(\ref{eq:dampingshift})
$b_\therm \approx  -{22}/{5} +11(1-n_s)/10$. Note that this approximation can in the future be improved by numerically recalibrating  the Green function $\mathcal{J}_\mu$ in the
separate universe so we parameterize the result as a 
``bias" factor $b_\therm$, similar to the slow-roll calculation of Ref.~\cite{Cabass:2018jgj} but with respect to the horizon scale at the end of thermalization rather than the end of the $\mu$ epoch.    For example if we assume that the functional form of $\mathcal{J}_\mu$ remains the same and only $z_\therm$ changes according to Eq.~(\ref{eq:athermL}), then 
$b_\therm \approx -4.1$ for $k_{\rm dip}=k_{\rm min}$.

Given the level of precision in these estimates, we simply take
\begin{equation}
b_\therm = -22/5
\end{equation}
as our fiducial bias.

\begin{figure}[t!]
\begin{center}
\includegraphics[width=1\columnwidth]{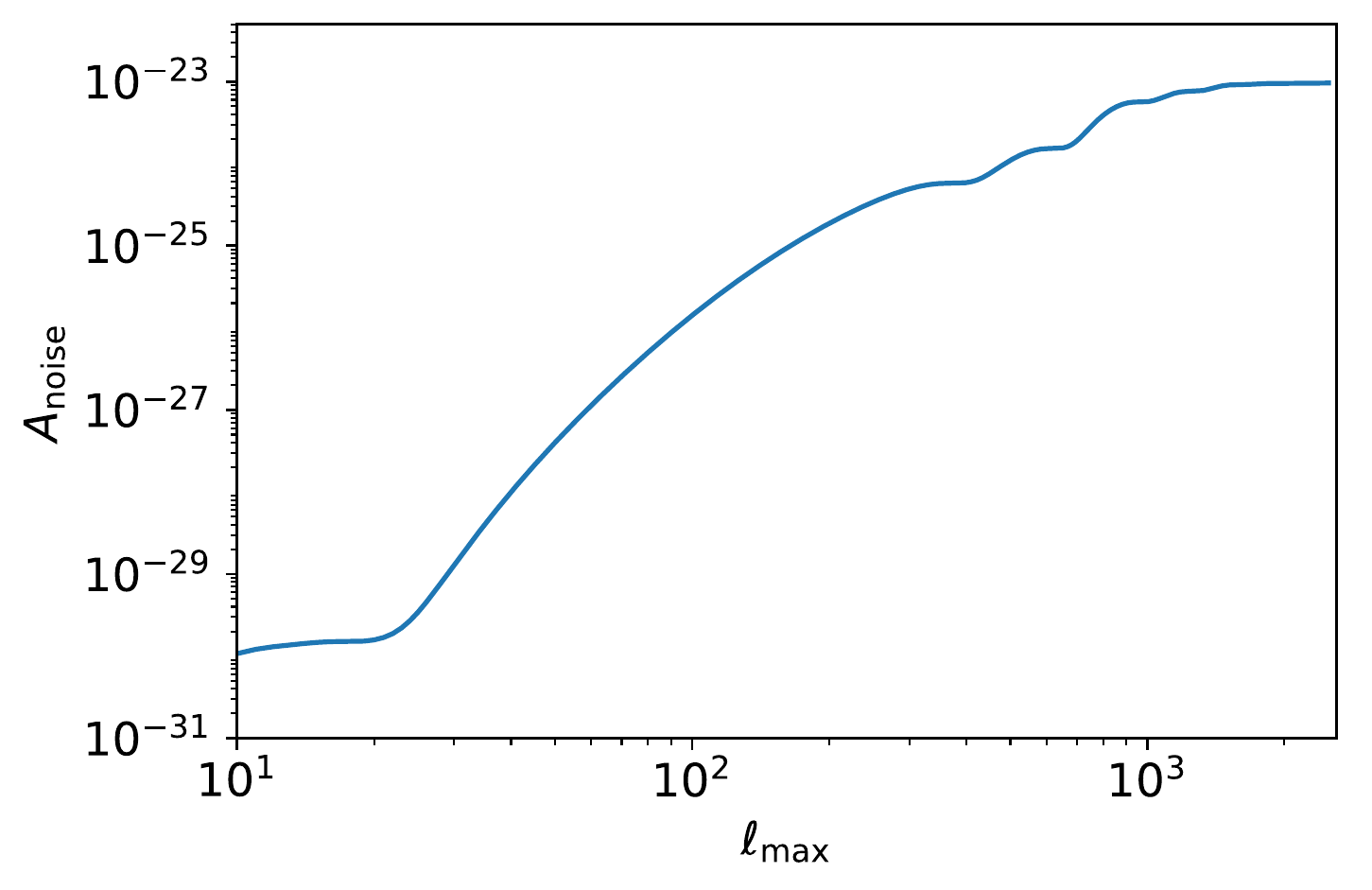}
\end{center}
\caption{\label{fig:noise}  Maximum $\mu$ white noise level $C_\ell^{\mu\mu}=A_{\rm noise}$ that allows for a $S/N=1$ measurement of the maximal $\bar \mu=9\times 10^{-5}$ and fiducial $b_\therm=-22/5$ PBH signal as a function of largest multipole measured $\ell_{\rm max}$.  $A_{\rm noise}$ saturates at $10^{-23}$ around $\ell_{\rm max}\sim 10^3$ and other models can be scaled as $A_{\rm noise} \propto b_\therm^2 \bar\mu^2$.
}
\end{figure}

Following Ref.~\cite{Cabass:2018jgj}, we can characterize the long wavelength correlation between CMB temperature anisotropy and $\mu$ anisotropy with the angular cross power spectrum
\begin{equation}
\label{eq:ClmuT}
    C_{\ell}^{\mu T} \equiv \bar\mu b_\therm C_{\ell}^{\mu T,b_\therm},
\end{equation}    
where
\begin{eqnarray}
C_{\ell}^{\mu T,b_\therm}
    =\frac{4\pi}{ (a H)_\therm^2} \int_{0}^{\infty} {d} k k \Delta^2_{\zeta}(k) \Delta_{\ell}^{\mu}(k) \Delta_{\ell}^{T}(k).
    \label{eq:ClmuTbmu}
\end{eqnarray}
Here $\Delta_\ell^T$ is the CMB temperature transfer function which we take from CAMB\footnote{
\href{https://camb.info}{https://camb.info}}
and
\begin{equation}
\Delta^\mu_\ell(k) = e^{-k^2/q_{\mu,D}^2 } j_\ell(\eta_0-\eta_{\rm rec}),
\end{equation}
where $q_{\mu,D} \approx 0.084$ Mpc$^{-1}$ 
 is the dissipation scale of $\mu$ inhomogeneities at recombination
\cite{Pajer:2012qep}.\footnote{This approximation can also be refined in the future with the Green function for spectral spatial anisotropy (Chluba, private communication).}  Notice that this damping factor is
comparable to that of the temperature transfer function
at $k_D(z_{\rm rec})$ 
and in combination they limit the integral in Eq.~(\ref{eq:ClmuT}) to the long wavelength $k_L$ values of the CMB.   It is straightforward to generalize this result for the cross correlation with CMB polarization with the polarization transfer function
which we leave that to a future work.

Finally notice that the primordial deviation from the dilation or consistency relation and the separate universe
growth of acoustic oscillations takes the 
same form as Eq.~(\ref{eq:ClmuT}) but are suppressed
by factors of $(a H/k_{\rm dip})^2_\therm \lesssim 4 \times 10^{-5}$ and
$(a H/k_D)^2_\therm \approx  10^{-7} $ respectively and hence provide a negligible correction to  $C_\ell^{\mu T}$
as calculated in Eq.~(\ref{eq:ClmuT})
from the thermalization bias.

\section{Signal-to-Noise}
\label{sec:snr}

The $\mu T$ cross power spectrum as calculated in the previous section is enhanced in PBH models by the $k^4$ rise in the small scale curvature power spectrum as long as $k_{\rm dip} \ll k_D(z_\therm)$ but also suppressed by the smallness of the density perturbation associated with 
$\zeta_L$ at the thermalization epoch $(k_L/aH)^2_\therm \zeta_L$.   Consequently unlike $\mu T$ correlations from nearly scale invariant perturbations in multifield inflation \cite{Pajer:2012vz,Ganc:2012ae,Pajer:2012qep,Chluba:2016aln,Cabass:2018jgj,Remazeilles:2021adt}, the signal-to-noise is dominated by the smallest angular scales at which the correlation can be detected, namely the damping scale of primary CMB anisotropy $k_L \sim 0.1$ Mpc$^{-1}$.   

To estimate the signal-to-noise in $\mu T$, we take the
Gaussian approximation for $\mu$ anisotropy 
\begin{equation}
\label{eq:snr}
\left( \frac{S}{N} \right)^2 = \sum_{\ell=2}^{\ell_{\rm max}} (2\ell+1)
\frac{ (C_\ell^{\mu T})^2 } { (C_\ell^{\mu T})^2 + C_\ell^{\mu\mu} C_\ell^{T T} },
\end{equation}
where the $C_\ell^{\mu\mu}$ and $C_\ell^{TT}$ terms in the denominator include both the sample variance of the signal and any instrumental noise from their measurement.

Given the smallness of the $\mu$ anisotropy, $C_\ell^{\mu\mu}$ will be noise dominated for the forseeable future.\footnote{The ultimate limit comes from the residual fluctuations from the averaging across CMB scales of the patches that dissipate \cite{Pajer:2012vz,Pajer:2012qep,Cabass:2018jgj} but note that in the PBH context the patches that contribute most to $\bar \mu$ are much smaller in size than the SR contributions at the end of the $\mu$ epoch and do not appreciably enhance the noise.}  To assess the maximal white noise level $C_\ell^{\mu\mu} = A_{\rm noise}$ at which the
signal is barely detectable at $S/N=1$, we can drop the
$C_\ell^{\mu T}$ sample variance term in the denominator of Eq.~(\ref{eq:snr}) and solve for white noise term
\begin{eqnarray}
{A_{\rm noise}} & =& 
 \sum_{\ell=2}^{\ell_{\rm max}} (2\ell+1)
\frac{ (C_\ell^{\mu T})^2 } { C_\ell^{T T} } \\
& = & {b_\therm^2 \bar \mu^2} \sum_{\ell=2}^{\ell_{\rm max}}  (2\ell+1)
\frac{ (C_\ell^{\mu T,b_\therm})^2 } { C_\ell^{T T} }. \nonumber
\end{eqnarray}
To estimate the maximal noise level for detection, we assume that the $TT$ measurement is cosmic variance limited to $\ell_{\rm max}$.   In Fig.~\ref{fig:noise}, we show this maximal noise amplitude as a function of $\ell_{\rm max}$ for the maximal signal  $\bar\mu=9 \times 10^{-5}$ and $b_\therm=-22/5$. 
Notice that the result saturates at around $\ell_{\rm max}\approx 10^3$  at a level of $\sim 10^{-23}$ since both $T$ and $\mu$ anisotropies are damped by diffusion.  
To constrain other PBH models, this result can be scaled as $A_{\rm noise} \propto {b_\therm^2 \langle \mu \rangle^2}$ using Eq.~(\ref{eq:mubarscaling}).

As this represents the maximal signal, detecting this effect will be challenging experimentally.   First, to optimize detection, an experiment would need at least several
arcminute scale resolution since
\begin{equation}
C^{\mu \mu}_\ell =
A_{\rm noise} e^{\frac{\ell^2 \theta_b^2}  {8 \ln 2}},
\end{equation} 
where $\theta_b$ is the full width half max of the beam in radians. 

Second, the required noise level  $A_{\rm noise} < 10^{-23}$ is quite stringent even ignoring foregrounds and systematics. 
For reference, a PIXIE-like mission \cite{Kogut:2011xw,Kogut:2020add}
which aims to measure $\overline{\mu}$ in single field slow roll
with $\theta_b \approx 1.6 ^{\circ}$,
will at best achieve $A_{\rm noise} \sim 10^{-15}$ 
which would suffice to constrain a squeezed bispectrum signal from multifield inflation to $\left|f_{\mathrm{NL}}\right| \lesssim 3000$ at 68\% CL \cite{Pajer:2012vz,Ganc:2012ae}.   To constrain the maximal PBH signal, one would require the equivalent sensitivity to $|f_{\rm NL}| < 0.3$ but with a much higher angular resolution.
LiteBIRD can potentially achieve a detector sensitivity of $A_{\rm noise} \approx 10^{-18}$ but still with only $\sim 0.5^\circ$ 
resolution and subject to foreground contamination \cite{Remazeilles:2021adt}.

\section{Discussion}
\label{sec:discussion}

We have shown that in spite of the large squeezed bispectrum due to the enhancement of small scale power in single field inflationary PBH models, the spectral distortion anisotropy is highly suppressed since for scales relevant to CMB cross correlation, it represents an unobservable modulation of global scales rather than of the local amplitude of the short wavelength modes.   Nonetheless the $\mu$ anisotropy can be larger than in single field slow roll inflation because of the large enhancement of
small scale power itself.   The largest effect comes from the local modulation of the expansion rate at a given locally measured efold from the end of inflation
and hence the end of the thermalization epoch.   This modulation provides a spatial variation in the amount of power in acoustic waves dissipated near thermalization that causes a $\mu$ distortion that is correlated with CMB anisotropy itself.   

This leading order correlation is enhanced by the $k^4$ rise in small scale power in PBH models but suppressed by the square of the ratio of
the comoving horizon at end of thermalization to observable CMB scales, 
$(k_L/aH)_\therm^2 \lesssim 0.0005$.   On the other hand, the enhancement of the average $\mu$ itself can be up to $\sim 5000$ and still satisfy current COBE-FIRAS bounds.   These compensating factors leave the signal potentially observable but still well beyond the capability of proposed space-based instruments like PIXIE and LiteBIRD.
Moreover to detect the correlation at the smallest observable CMB scales, where the signal peaks, would require a telescope with several arcminute scale resolution or better. 

These properties of the PBH $\mu T$ correlation suggest that in the future, ground based instruments may provide a competitive path forward, given the rapid advance in the scale of detectors deployed on large telescopes into the CMB-S4 \cite{CMB-S4:2016ple} era and beyond.  Unlike the absolute measurement of $\bar \mu$, systematics and foregrounds can also be mitigated by differential measurements and cross correlation \cite{Ganc:2012ae}.  Furthermore the $\mu T$ correlation can be supplemented by polarization cross correlation \cite{Ravenni:2017lgw, Remazeilles:2021adt}.  We leave these studies to a future work.

\acknowledgments

We thank Giovanni Cabass, Jens Chluba, Tom Crawford, Austin Joyce, Hayden Lee, Enrico Pajer, Sam Passaglia, and LianTao Wang for useful discussions.  D.Z.\
was supported by the National Science Foundation Graduate Research Fellowship Program under Grant No. DGE1746045. 
K.I.\ was  supported by the Kavli Institute for Cosmological Physics at the University of Chicago through an endowment from the Kavli Foundation and its founder Fred Kavli. W.H.\ 
was supported by U.S.\ Dept.\ of Energy contract DE-FG02-13ER41958 and the Simons Foundation. 
Portions of this work was performed at the Aspen Center for Physics, which is supported by National Science Foundation grant PHY-1607611.
\begin{widetext}
\appendix

\section{Dilation Consistency Relation at Second Order}
\label{sec:secondorder}

In this Appendix, we show that at second order in Newtonian gauge, the response of the short wavelength density perturbation to the long wavelength $\zeta_L$ is a pure dilation and when combined with the Maldacena consistency relation there is no locally observable effect of a constant $\zeta_L$.
Since we focus on the $\mu$-distortion, mainly produced during the radiation dominated epoch, we concretely calculate the second-order perturbations in that era.
Note that the calculation and notation here, which differs from the main text, 
is based on Ref.~\cite{Inomata:2020cck}.

First, let us summarize our notation in this appendix.
In Newtonian gauge, we can write the scalar parts of the metric perturbations as
\begin{align}
 d s^2 &= g_{\mu\nu} d x^\mu d x^\nu 
= a^2 \left\{- (1 + 2 \Phi^{(1)} + \Phi^{(2)})d \eta^2  
+ \left[ (1-2 \Psi^{(1)} - \Psi^{(2)}) \delta_{ij}\right] {d y^i d y^j }\right\},
\label{eq:def_metric_pertb}
\end{align}
where the superscript denotes the order in perturbations.
Here, we assume a perfect fluid for simplicity, which leads to $\Phi^\fo = \Psi^\fo$, and thus ignore the correction from neutrinos. 
Then, we can express the energy-momentum tensor as
\begin{align}
	T^\mu_{\ \nu} &=  (\rho + P) u^\mu u_\nu + P \delta^\mu_{\ \nu},
\end{align}
where $\rho$ is the energy density, $P$ is the pressure, and $u_\mu$ is the 4-velocity.
We take the following notation for their perturbations up to the second order:
\begin{align}
	\label{eq:rho_pertb_def}
	\rho &= \rho^\zo + \delta \rho^\fo + \frac{1}{2} \delta \rho^\so, \\
	P &= P^\zo + \delta P^\fo + \frac{1}{2} \delta P^\so, \\
	\label{eq:u_i_pertb}
	u^i &= \frac{1}{a} \left( \delta v^{\fo,i} + \frac{1}{2} \delta v^{\so,i} + \frac{1}{2} \delta {v^i_V}^\so \right),
\end{align}
where $\delta v$ is the velocity potential and $\delta v_V^i$ is the vector part of the velocity perturbation.
We define the density perturbation, $\delta^{(a)}$, as 
\begin{align}
	\delta^{(a)} \equiv \frac{\delta \rho^{(a)}}{\rho^{(0)}}.
\end{align}
At the first order in perturbations, this is related to the curvature perturbation as 
\begin{align}	
	\label{eq:phi_zeta_rel}
	\delta^\fo_{\bm k}(\eta) &= \displaystyle -\frac{2}{3} \zeta^\fo_{\bm k} T_{\delta}(x),
	 \end{align}
where the subscript $\bm k$ represents the Fourier mode whereas perturbations are in real space otherwise in this Appendix, $x\equiv k \eta$ with $k \equiv |\bm k|$, and $\zeta^\fo = -(3/2)\Phi^\fo$ in the superhorizon limit.
The transfer function is defined as 
\begin{align}
	T_{\delta}(x) &\equiv \frac{6x(-6 + x^2) \cos(\tfrac{x}{\sqrt{3}}) - 12 \sqrt{3} (-3 + x^2) \sin(\tfrac{x}{\sqrt{3}})}{x^3}.
\end{align}

Next, let us calculate the second-order density perturbations.
The density perturbation is related to the other perturbations as~\cite{Inomata:2020cck}
\begin{align}
	\delta^\so	=& -2 \Psi^\so +2 N^j_{\ i} {B^i_{\ j}}^\so -\frac{2}{\mathcal H} {\Psi^\so}' + \frac{2}{3\mathcal H^2} \Psi^{\so,i}_{\quad \  ,i} 
	+ \frac{2}{\mathcal H^2} \left({\Phi^\fo}' \right)^2 + \frac{16}{3\mathcal H^2} \Phi^\fo \Phi^{\fo ,i}_{\quad \ ,i}  \nonumber \\
	& + \frac{1}{\mathcal H^2} \left( 2 - \frac{8}{9(1+w)} \right) \Phi^{\fo ,i} \Phi^\fo_{\quad ,i} - \frac{8}{9(1+w) \mathcal H^3} (\Phi^{\fo ,i} \Phi^\fo_{\quad ,i} )'  - \frac{8}{9(1+w) \mathcal H^4} {\Phi^\fo}'_{, i}  {\Phi^{\fo,i}}',\label{eq:delta_so_exp}
\end{align}
where the prime here denotes the derivative with respect to $\eta$, $\mathcal H (\equiv a'/a)$ is the conformal Hubble parameter, and $w$ is the equation of state parameter.
$B^i_{\ j}$ and $N^i_{\ j}$ are defined as 
\begin{align}
 {B^i_{\ j}}^\so \equiv \left[\frac{4(5+3w)}{3(1+w)} \Phi^{\fo ,i} \Phi^\fo_{\quad ,j} 
+ \frac{8}{3(1+w) \mathcal H} \left(\Phi^{\fo ,i} {\Phi^\fo_{\quad ,j}}\right)' 
 + \frac{8}{3(1+w) \mathcal H^2} {\Phi^{\fo ,i}}' {\Phi^\fo_{\quad ,j}}'\right],
 \label{eq:b_i_j_def}
\end{align}
\begin{align}
	N^j_{\ i} A^i_{\ j}(\bm x) &\equiv \frac{3}{2} \nabla^{-2} \left( \frac{\partial^j \partial_i}{\nabla^{2}} - \frac{1}{3} \delta^j_{\ i} \right) A^i_{\ j} (\bm x) \nonumber \\
	&= \int \frac{d^3 k}{(2\pi)^3} \left( -\frac{3}{2k^2} \right) \left( \frac{k^j k_i}{k^2} - \frac{1}{3} \delta^j_{\ i} \right) {A_{\bm k}}^i_{\ j} , \label{eq:n_ji_def}
\end{align}
where $A^i_{\ j}$ is an arbitrary tensor.
Since $\delta^\so$ depends on $\Psi^\so$, we need to calculate $\Psi^\so$ first. 
The equation of motion for $\Psi^\so$ is~\cite{Inomata:2020cck}
\begin{align}
{\Psi^\so}'' + 3 (1 + c_s^2) \mathcal H {\Psi^\so}' + \left[2\mathcal H' + (3c_s^2 + 1) \mathcal H^2 \right] \Psi^\so - c_s^2 \Psi^{\so,i}_{\quad \  ,i} = S^\so, 
	\label{eq:psi_so_eom_re}
	\end{align}
\begin{align}
	\label{eq:psi_so_source_re}
	S^\so \equiv& \left(3c_s^2 -\frac{1}{3} \right) \Phi^{\fo ,i} \Phi^\fo_{\quad ,i} + 8c_s^2 \Phi^\fo \Phi^{\fo ,i}_{\quad \ ,i} + (3c_s^2 + 1) \left({\Phi^\fo}' \right)^2 + \left[ (3c_s^2 + 1)\mathcal H^2 + 2 \mathcal H' \right]  N^j_{\ i} {B^i_{\ j}}^\so \nonumber \\
	& \ 
	 + \mathcal H  N^j_{\ i} \left({B^i_{\ j}}^\so \right)' + \frac{1}{3}  N^j_{\ i} \left({B^i_{\ j}}^\so \right)^{,k}_{\ ,k} + \left(\frac{1}{3} - c_s^2 \right) \frac{4}{3(1+w) \mathcal H^2}  \left( \mathcal H \Phi^{\fo,i} +  {\Phi^{\fo,i}}' \right)  \left(  \mathcal H \Phi^\fo_{,i} + {\Phi^\fo}'_{, i} \right),
\end{align}
where $c_s$ is the sound speed.

Here, we focus on the second-order perturbations induced by a superhorizon mode ``$L$'' on the much smaller mode $k_L/k\ll 1$.
This case corresponds to the $\mu$-distortion production in the presence of the superhorizon perturbations.
In this case, we can approximate Eq.~(\ref{eq:psi_so_eom_re}) as 
\begin{align}
 {\Psi^\so}'' + 4 \mathcal H {\Psi^\so}'  - \frac{1}{3} \Psi^{\so,i}_{\quad \  ,i} \simeq \frac{8}{3} \Phi^\fo_L \Phi^{\fo ,i}_{\quad \ ,i},
 \label{eq:psi_eom}
\end{align}
where we have substituted 
The other contributions in $S^\so$ are sub-leading because they include spatial derivative on the superhorizon mode ``$L$''.
In the Fourier space, the equation becomes 
\begin{align}
	{\Psi^\so}''_{\bm k} + \frac{4}{\eta} {\Psi^\so}'_{\bm k} + \frac{k^2}{3} \Psi_{\bm k}^\so \simeq -\frac{8}{3} \Phi_L^\fo k^2 \Phi_{\bm k}^\fo,
\end{align}
where we have assumed that the superhorizon perturbation as the constant quantity, $\Phi_L$, because it does not evolve in the Newtonian gauge.
Note that $\Phi_L$ here and below is still a function of spatial comoving coordinates but variations are small locally compared to the short wavelength mode so we have ignored them in the Fourier transform.
Then, we can solve $\Psi^{(2)}_{\bm k}$ using the Green function method:
\begin{align}
	\Psi^\so_{\bm k} (\eta) &\simeq \Psi^\so_{\bm k}(0) T(x) + \int^\eta_0 d \bar\eta \left(\frac{a(\bar\eta)}{a(\eta)}\right)^2 G(k,\eta;\bar \eta) \left( -\frac{8}{3} \Phi_L k^2 \Phi^\fo_{\bm k}(\bar \eta) \right),
\end{align}
where the concrete expression of the Green function is given by 
\begin{align}
	k G (k, \eta;\bar \eta) = -\Theta(\eta - \bar \eta) \frac{x \bar x}{\sqrt{3}} \left[ j_1(x/\sqrt{3}) y_1(\bar x/\sqrt{3}) - j_1(\bar x/\sqrt{3}) y_1(x/\sqrt{3}) \right].
\end{align}
The initial condition of $\Psi^\so$ is given by
\begin{align}
	\Psi^\so_{\bm k}(0) = -\frac{16}{9} \zeta^\fo_L \zeta^\fo_{\bm k},
	\label{eq:psi_so_r}
\end{align}
where we have here assumed a Gaussian distribution of $\zeta$, which means that the contribution from primordial non-Gaussianity from the Maldacena consistency relation is not included in $\Psi^\so$ (and $\delta^\so$) in this expression. 
We will independently take into account that contribution later.
Then, we obtain $\Psi^\so$ as 
\begin{align}
	\Psi^\so_{\bm k}(\eta) 
	& \simeq \left[\frac{16\sqrt{3}}{3} \frac{\sin(x/\sqrt{3})}{x} + \frac{64\cos(x/\sqrt{3})}{x^2}\right] \zeta_L^\fo \zeta^\fo_{\bm k}, \label{eq:psi_second_sq2}
\end{align}
where we have neglected the contribution of $\mathcal O(x^{-3})$ because we want to know the evolution of the second-order density perturbations on  subhorizon scales, which is related to the $\mu$-distortion production.
In the large $x$ limit, Eq.~(\ref{eq:delta_so_exp}) can be approximated in the Fourier space as 
\begin{align}
	\delta^\so_{\bm k} &\simeq -\frac{2x^2}{3} \Psi^\so_{\bm k} - 2x \frac{d \Psi^\so_{\bm k}}{d x}  - \frac{16x^2}{3} \Phi_L^\fo \Phi^\fo_{\bm k} \nonumber \\
	& \simeq  \left(-\frac{32\sqrt{3}}{9} x \sin(x/\sqrt{3}) - 32 \cos(x/\sqrt{3}) \right)  \zeta_L^\fo \zeta^\fo_{\bm k}.
\end{align}
Then, we finally get the following expression for the total energy density perturbation:
\begin{align}
	\delta_{\bm k}(\eta) &= \delta^\fo_{\bm k}(\eta) + \frac{1}{2} \delta^\so_{\bm k}(\eta) \nonumber \\
	&\simeq -\frac{2}{3} \left( 1 + \frac{4}{3}\zeta^\fo_L \right) T_\delta\left(x \left(1 - \frac{4}{3} \zeta^\fo_L \right) \right) \zeta^\fo_{\bm k},
	\label{eq:delta_rad_so}
\end{align}
where the approximate equality is valid in $x \gg 1$.

In the following, we relate this result to the dilation transformation. 
The dilation consistency relation in the Newtonian gauge is given by~\cite{Horn:2014rta}
\begin{align}
  \lim_{\bm q \rightarrow 0} \frac{\vev{\pi_{\bm q} \delta_{\bm k_1} \cdots \delta_{\bm k_N}}^c}{P_\pi(q)} \epsilon(\eta) = \sum^N_{a=1} \left[ -\epsilon(\eta_a) \left( \left.\frac{\bar \rho'}{\bar \rho}\right|_{\eta_a} + \partial_{\eta_a} \right) + \lambda \left(3 + \bm k_a \cdot \partial_{\bm k_a} \right) \right] \vev{\delta_{\bm k_1} \cdots \delta_{\bm k_N}}^c,
  \label{eq:cons_rel}
\end{align}
where $\vev {\pi_{\bm q} \pi_{\bm p}} = (2\pi^3) \delta_D(\bm q + \bm p) P_\pi(q)$, $\pi$ is the velocity potential ($\pi = \delta v$), and the superscript $c$ of the braket means the connected part.
The $\epsilon$ and $\lambda$ are the coordinate transformation parameters to go from a gauge where the effect is a pure dilation and are given by 
  {$y^\mu \rightarrow {y'}^\mu = y^\mu + \xi^\mu$ with $\xi^0 = \epsilon$ and $\xi^i = \lambda y^i$.}
During a radiation dominated epoch, we can derive the following relations: $\lambda = -3\epsilon/\eta$, $\bar \rho'/\bar \rho = -4/\eta$, and $\pi \simeq \eta \zeta/3$ in the superhorizon limit.
Here, we take $\lambda = \zeta_L$ and $N=2$ to determine the modulation of short wavelength power by $\zeta_L$, such that the transformation is from comoving gauge to Newtonian gauge
and rewrite the consistency relation as 
\begin{align}
  \lim_{\bm q \rightarrow 0} \frac{\vev{\zeta_{\bm q} \delta_{\bm k_1} \delta_{\bm k_2}}^c}{P_\zeta(q)} \zeta_L^\fo = \sum^2_{a=1} \left[ \left( \frac{4}{3} \zeta_L^\fo - \frac{1}{3} \zeta_L^\fo \eta\, \partial_{\eta} \right) - \zeta_L^\fo \left(3 + \bm k_a \cdot \partial_{\bm k_a} \right) \right] \vev{\delta_{\bm k_1} \delta_{\bm k_2}}^c.
  \label{eq:cons_rel2}
\end{align}
{Note again that $\zeta_L$ is in real space as $\zeta_L(\bm y)$, but its perturbation scale is much larger than the scales of $\bm k_1$ and $\bm k_2$ so that we can approximately consider it constant except when considering the correlation to the large scale mode} $\bm q$.
Using Eq.~(\ref{eq:delta_rad_so}), we can rewrite the left-hand side as 
\begin{align}
  \lim_{\bm q \rightarrow 0} \frac{\vev{\zeta_{\bm q} \delta_{\bm k_1} \delta_{\bm k_2}}^c}{P_\zeta(q)} \zeta_L^\fo  =& \frac{4}{9} \left[\left(1+ \frac{8}{3}\zeta^\fo_L \right) T_\delta^2 \left(x_1 \left(1 - \frac{4}{3} \zeta^\fo_L \right) \right) - T_\delta^2 \left(x_1  \right)\right] (2\pi^3) \delta_D(\bm k_1 + \bm k_2) P_\zeta(k_1) \nonumber \\
  &- \zeta_L^\fo
  \frac{d\ln \Delta_\zeta^2}{d\ln k} \Big|_{k=k_1}
  (2\pi^3) \delta_D(\bm k_1 + \bm k_2) P_\zeta(k_1),
  \label{eq:3pt_func_cons_2}
\end{align}
where we have used 
\begin{align}
    \vev{\zeta_{\bm q}^\fo \zeta_L^\fo(\bm y)}^c &= \int \frac{d^3 k}{(2\pi)^3} e^{i \bm k \cdot \bm y} \vev{\zeta^\fo_{\bm q} \zeta^\fo_{\bm k}}^c \nonumber \\
    &= P_\zeta(q).
\end{align}
The first line in Eq.~(\ref{eq:3pt_func_cons_2}) corresponds to the contribution from $\delta^\so$. On the other hand, the second line corresponds to the Maldacena consistency relation~\cite{Maldacena:2002vr}, that is, it just comes from the contribution proportional to $\vev{\zeta_{\bm q} \zeta_{\bm k_1} \zeta_{\bm k_2}}$, not related to the evolution of the second-order scalar perturbations.
On the other hand, once we substitute 
\begin{align}
  \vev{\delta_{\bm k_1} \delta_{\bm k_2}}^c = \frac{4}{9} T_\delta^2(x_1) (2\pi^3) \delta_D(\bm k_1 + \bm k_2) P_\zeta(k_1)
\end{align}
into the right-hand side of Eq.~(\ref{eq:cons_rel2}), we can see that the right-hand side is the same as Eq.~(\ref{eq:3pt_func_cons_2}) at least at the lowest order in the perturbations, $\mathcal O(\zeta_L P_\zeta)$.
From this, we can see that the dilation consistency relation is satisfied, once both the second-order evolution and primordial non-Gaussianity of the Maldacena relation are included, which indicates that the constant $\zeta_L$ does not give locally observable effects.
Although our calculation is based on the perfect fluid assumption, we can expect that the dilation consistency relation would be satisfied even in imperfect fluid.
\end{widetext}
\bibliography{muTPBH}
\end{document}